# THERMAL RECOVERY OF COLOUR CENTRES INDUCED IN CUBIC YTTRIA-STABILIZED ZIRCONIA BY CHARGED PARTICLE IRRADIATIONS

**Jean-Marc COSTANTINI**[*], CEA/SACLAY/DMN/SRMA, F-91191 Gif-Sur-Yvette Cedex France,

**François BEUNEU**, LSI, CEA-CNRS-Ecole Polytechnique, F-91128 Palaiseau Cedex, France.

**ABSTRACT**

We have used electron paramagnetic resonance to study the thermal annealing of colour centres induced in cubic yttria-stabilized zirconia by swift electron and heavy ion-irradiations. Single crystals were irradiated with 1 or 2-MeV electrons, and 200-MeV $^{127}$I, or 200-MeV $^{197}$Au ions. Electron and ion beams produce the same colour centres: namely i) an $F^+$-like centre, ii) the so-called T-centre ($Zr^{3+}$ in a trigonal oxygen local environment), and iii) a hole center. Isochronal annealing was performed up to 973 K. Isothermal annealing was performed at various temperatures on samples irradiated with 2-MeV electrons. The stability of paramagnetic centres increases with fluence and with a TCR treatment at 1373 K under vacuum prior to the irradiations.

Two distinct recovery processes are observed depending on fluence and/or thermal treatment. The single-stage type I process occurs for $F^+$-like centres at low fluences in as-received samples, and is probably linked to electron-hole recombination. T-centres are also annealed according to a single-stage process regardless of fluence. The annealing curves allow one to obtain activation energies for recovery. The two-stage type II process is observed only for the $F^+$-like centres in as-received samples, at higher fluences, or in reduced samples. These centres are first annealed in a first stage below 550 K, like in type I, then transform into new paramagnetic centres in a second stage above 550 K. A simple kinetics model is proposed for this process. Complete colour centre bleaching is achieved at about 1000 K.

**PACS numbers:** 61.80.Fe, 61.80.Jh, 61.82.Ms, 76.30.Mi

**Keywords: zirconia, colour centres, electron and ion irradiations.**

---

[*] Corresponding author : jean-marc.costantini@cea.fr

# I. INTRODUCTION

Zirconia (ZrO$_2$) is one of the most studied refractory materials suitable as a host for actinide transmutation [1]. Therefore, the knowledge of the stability of defects induced under ionizing and/or displacive radiations is a very important issue. On the basis of electron paramagnetic resonance (EPR) and optical absorption spectroscopy data [2-5], we have shown that two colour centres are produced in cubic yttria-stabilized zirconia or YSZ, after irradiations with swift electron or heavy ion beams: i) the first one is identified as an F$^+$-like centre (singly ionized oxygen vacancy, V$_O^{\cdot}$) with a small g-factor anisotropy (g$_\perp$=1.972 and g$_{//}$=1.996), and oriented along a <100> direction, most probably with one nearest-neighbour (NN) doubly-ionized oxygen vacancy (V$_O^{\cdot\cdot}$), ii) the second one is the so-called T-centre which is also produced by UV or X-ray irradiations, or thermo-chemical reduction (TCR) at high temperature under vacuum [6], i.e. Zr$^{3+}$ sitting in a site with a trigonal (C$_{2v}$) symmetry with two NN doubly-ionized oxygen vacancies (V$_O^{\cdot\cdot}$), oriented along a <111> direction, and with a large g-factor anisotropy (g$_\perp$=1.855 and g$_{//}$=1.986). The F$^+$- like centre was associated with a broad optical absorption band centred at a wavelength near 500 nm, giving a light-purple colouration, with an absorption coefficient proportional to the room-temperature (RT) volume density of centres deduced from the EPR spectra [2]. It was previously shown that T-centres give rise to an absorption band centred at 375 nm, giving a yellow colour [6]. The yield of F$^+$-type centres is scaled with the atomic displacements induced by elastic collisions, whereas T-centres are most probably induced by the ionization and electronic excitation processes [2].

Previous results with 2.5-MeV electron and 100-MeV $^{13}$C irradiations showed evidence of two distinct recovery processes [3, 4]. The first one is a single-stage annealing process of both colour centres which was modelled by first-order reaction kinetics [4]. The second one is a two-stage process. In the second stage, F$^+$- like centres were found to transform into new paramagnetic centres above 600 K either in the case of high-fluence irradiations or of TCR at 1373 K under vacuum prior to the irradiations [4]. The aim of the present paper is then i) to bring new data on the thermal recovery of these colour centres in YSZ single crystals after irradiation with lower energy electrons or much heavier ions, ii) to provide a deeper analysis of the second recovery process.

# II. EXPERIMENTAL PROCEDURE

YSZ single crystals, i.e. ZrO$_2$:Y (with 9.5 mol% Y$_2$O$_3$, ρ=5.8 g cm$^{-3}$), with the (100) orientation and thickness of 500 µm, provided by Crystal-GmbH (Berlin, Germany), were used in this study. The transparent and colourless as-received samples with formula $Zr_{0.8}Y_{0.2}O_{1.9}$ contain around 10 at% doubly ionized oxygen vacancies with a 2+ charge state induced by charge compensation of $Y^{3+}$ substituted for $Zr^{4+}$ ions.

The 1.0-MeV and 2.0-MeV electron irradiations were performed at the Van de Graaff accelerator of the LSI (Palaiseau) under helium gas at around 310 K with a flux around $10^{14}$ $cm^{-2} s^{-1}$. The electrons were transmitted through the samples. Ion irradiations were carried out at the VIVITRON (Strasbourg) with 200-MeV $^{127}I$ and 200-MeV $^{197}Au$ ions. Ions were implanted at various fluences in the samples under high vacuum with low fluxes ($\approx 10^8$-$10^{10}$ $cm^{-2} s^{-1}$) in order to prevent sample heating or charging. The temperature during irradiations was thus near RT. Irradiation characteristics of samples are given in Tables 1-2. Samples irradiated with 2.0-MeV electrons were kept at 77 K after irradiation.

Isochronal thermal annealing was carried out in air in an alumina crucible for 30 min from 323 K to 973 K on samples irradiated at low fluences, in the range of linear increase of defect density versus fluence, or at higher fluences (about 10 times larger), in the saturation regime, depending on the projectile and its energy [2]. Isothermal annealing in air versus time was also performed at 350 K, 368 K and 394 K on samples irradiated with 2.0-MeV electrons at the same fluence ($3.2 \times 10^{18}$ $cm^{-2}$). The defect recovery was monitored by RT EPR measurements with B//<100> using a computerized Bruker ER 200D X-band spectrometer operating at 9.6 GHz. The estimated uncertainty on the g-factor values measured by NMR is ± 0.001.

### III. RESULTS

Fig. 1(a) displays the RT EPR spectra with B//<100> of an YSZ sample irradiated with 2.0-MeV electrons at a fluence of $3.2 \times 10^{18}$ $cm^{-2}$ (sample q). No signal was found in this magnetic field range for pristine as-received samples. The same three main lines are seen with B//<100> regardless of fluence, either with electrons or ion irradiations [2]: i.e. a narrow line at g=1.998 (line 1), a broad signal at g=1.973 (line 2) and a very broad one at g=1.904 (line 3). Lines 1-2 have been previously assigned [2] to an $F^+$-type centre and line 3 to the so-called T-centre [2-4], also produced by photon irradiations [6]. The latter line was weak in the case of I or Au ion-irradiated as-received samples, like in previous results with C ions [2, 4].

Three additional weak lines are seen at g=2.014 (line 4) and g=2.007 (line 5), corresponding to a hole centre (Fig. 1(a)), and a T-centre satellite at g=1.97, similar to that of the TCR-induced centre [5-6]. At low fluences, where the number of $F^+$-type centres is small, the latter satellite line can be confused with line 2 at about the same g-factor value. However, at large electron fluences or energies, these three extra lines become very weak with respect to lines 1 and 2 [2-5].

EPR lines were simulated with Lorentzian curves and integrated intensities deduced from a least-square fitting procedure. The number of paramagnetic centres was then deduced from a calibration using a copper sulfate reference sample. The volume densities (N) were then obtained by dividing the latter values by the irradiated sample volume, i.e. surface times the thickness for electrons, or the mean projected range for ions. Densities were normalized to the starting values ($N_0$) (Tables 1-2) prior to annealing, and plotted versus annealing temperature in the isochronal annealing curves (Figs. 2-3). Isothermal annealing curves for the 2.0-MeV electron irradiated samples (q, r, t) were obtained by plotting $N/N_0$ versus annealing time (Fig. 4). Two distinct (type I and type II) recovery processes were found in the isochronal annealing curves.

Type I is a single-stage annealing process occurring for $F^+$-like centres in ion-implanted (I-14, I-18, AU-5, AU-11) (Fig. 3), and electron-irradiated samples (j, s) at low defect densities (Fig. 2) in the range of linear increase versus fluence, and also for T-centres in all electron-irradiated samples (e, j, s) (Fig. 2). $F^+$-type centres were bleached near 500-550 K in all samples. In the sample (e) irradiated with 1.0-MeV electrons at low fluence, the $F^+$-like centres vanished rapidly at RT, so that it was impossible to measure the annealing curves. In the latter sample (e), the T-centres were bleached at about 600 K like the X-ray induced T-centres [6], whereas at high fluences in the saturation regime (sample j), they were bleached near 1000 K, like the TCR-induced T-centres [6] (Fig. 2). These results are in agreement with our previous data for both centres induced by 2.5-MeV electron or 100-MeV C ion irradiations [4]. In the present ion-irradiated samples, however, it was not possible to monitor the T-centre annealing since EPR signals were too weak.

During annealing, the electron-irradiated samples first turned from a brown or gray colour to light-yellow, corresponding to the 375-nm absorption band of T-centres, and were progressively bleached. The ion implanted samples turned from a light-purple colour, corresponding to the 500-nm absorption band of $F^+$-like centres, to light-yellow, and were also subsequently bleached. The two lines 4 and 5, probably corresponding to oxygen ($O^-$) hole centres (g>2.0023) [6], were bleached at 330 K and faded out within few hours at RT in the 1.0-MeV electron irradiated samples (e, j), like in X-ray irradiated samples [6]. However, these hole centres were found to be more stable up to 473 K in the 2.0-MeV electron irradiated samples (q, r, s, t) (Fig. 2). They followed the same isochronal and isothermal annealing behaviour as the $F^+$-like centres (Figs. 2, 4). The T-centre satellite line at g=1.97 followed a similar isochronal annealing behaviour as line 3 (Fig. 2).

Type II is a more complex two-stage process which was previously observed for the $F^+$-like centres in 2.5-MeV electron and 100-MeV C ion irradiated as-received samples at higher fluences, in the saturation regime, with larger defect densities [4]. It was also found in reduced samples annealed at 1373 K under vacuum prior to the irradiations, regardless of fluence [4]. In this second recovery process: i) the first stage, below 550 K, was similar to type I, ii) the second one, above 550 K, exhibited modifications of lines 1 and 2. In this second stage, a new line (line 6) appeared (Fig. 1(b)) and caused apparent increase of $F^+$-type centres [4]. Accordingly, the samples turned to a maroon colouration indicating the formation of new colour centres at about 600 K. Complete colour centre bleaching is achieved at about 1000 K [4]. Such a process was not seen in the present data for lower-energy electron or heavier ion irradiations.

### IV. DISCUSSION
### IV.1. Type I recovery process

This process is associated to the sigmoid isochronal annealing curves (Figs. 2-3) either for $F^+$-like or T centres. These sets of curves are first analyzed on the basis of first-order reaction kinetics, as was done for the X-ray induced T-centres [9], with the following rate equation:

$$dN/dt = -\nu N \qquad (1).$$

Using the assumption that the frequency factor ($\nu$) is thermally activated:

$$\nu = \nu_0 \exp(-\Delta\varepsilon/k_B T) \qquad (2),$$

gives a classical thermally-activated behaviour for the annealing curve:

$$N/N_0 = \exp[-\nu_0 t_a \exp(-\Delta\varepsilon/k_B T)] \qquad (3),$$

where $\nu_0$ is a frequency factor, $t_a$=1800 s, the annealing time, $k_B$, the Boltzmann constant, T, the annealing temperature, and $\Delta\varepsilon$, the activation energy for recovery.

Least-square fitted parameters ($\Delta\varepsilon$, $\nu_0$) deduced from Eq(3) are displayed in Table 1 for $F^+$-like centres, and Table 2 for T-centres (samples e, j, s, I-6, I-14, I-18, AU-5, AU-11). For each sample, the activation energies of line 1 and line 2 are in good agreement (Table 1), thereby confirming that these two lines correspond to one and the same defect [2]. The corresponding fitted annealing curves are plotted in Figs. 2-3. Activation energies are on the whole compatible with the calculated activation energy for migration of oxygen vacancies in YSZ of 0.73 eV [10]. An activation energy of 0.5 eV and $\nu_0=100$ s$^{-1}$ were obtained for the annealing of the X-ray induced T-centres in YSZ (with 12 and 24 mol% $Y_2O_3$) [9], in agreement with our data (Table 2). However, first-order reaction kinetics should imply in that case that atomic point defects are annihilated at fixed sinks [11]. An alternative explanation could involve thermal activation of electrons from deep donor levels, corresponding to these electron centres, to the conduction band (band gap energy: $E_g$=4.23 eV [12]), or to other localized levels in the band gap.

We have also deduced from the fitted ($\Delta\varepsilon$, $\nu_0$) values the characteristic temperatures ($T_{1/2}$) (Tables 1-2), corresponding to the curve inflection point for a first-order kinetics, at which half of centres are annealed out:

$$T_{1/2} = \Delta\varepsilon/\{k_B [\ln(\nu_0 t_a)-\ln(\ln 2)]\} \approx \Delta\varepsilon/[k_B \ln(\nu_0 t_a)] \qquad (4).$$

For F$^+$-like centres in ion-irradiated samples, $T_{1/2}$ is about 450-550 K, and increases versus fluence like for C ion irradiations [4]. However, a clear difference was found with the 2.5-MeV electron irradiated samples at the same high fluence (1x10$^{19}$ cm$^{-2}$), in which F$^+$-like centre recovery occurred according to type II [4]. This is most probably due, in the latter case, to the larger T-centre density ($N_0 \approx 6\times10^{18}$ cm$^{-3}$) [4] with respect to the 1.0-MeV electron irradiation (sample j: $N_0 \approx 2\times10^{18}$ cm$^{-3}$) (see below). For T-centres in the electron-irradiated samples, $T_{1/2}$ increases from 480 K (sample e) to about 870 K (sample j), when increasing fluence by a factor about 10, from the linear regime to the saturation of defect production, as previously found [4]. Note that F$^+$-like and T centres are always more stable at larger fluences, as shown previously [3-5].

Within this assumption, the time-decay curve for isothermal annealing should be then:

$$N/N_0 = \exp(-t/\tau) \qquad (5),$$

with:

$$\tau = \nu_0^{-1} \exp(\Delta\varepsilon/k_B T_a) \qquad (6),$$

where $\tau$ is the defect life time at a given annealing temperature $T_a$. Analyses of the isothermal annealing curves of F$^+$-like centres in 2.0-MeV irradiated samples at 394 K (q), 368 K (r) and 350 K (t) shows large deviations from the exponential decay approximation at long times on the plot of $N/N_0$ versus reduced annealing time ($t/\tau$) (Fig. 4), where $\tau$ is deduced from the decay at short time scales ($t<2\tau$) (Table 1). A similar discrepancy is found for line 5 (hole centres) in sample q (at 394 K) also exhibiting the same asymptotic behaviour at longer times (Fig. 4). However, isochronal annealing curve of F$^+$-like centres in sample s with the same $N_0$ value as these samples exhibits a single recovery stage (Fig. 2) with $\Delta\varepsilon=0.59$ eV and $T_{1/2}\approx 490$ K compatible with a first-order kinetics (Table 1).

We have thus used higher-order ($\gamma >1$) reaction kinetics with the following rate equation:

$$dN/dt = -\nu' N^\gamma \qquad (7).$$

Using Eq(2) yields after integration for the isochronal annealing curves:

$$N/N_0 = [1 + \nu_0' N_0^{\gamma-1} (\gamma-1) t_a \exp(-\Delta\varepsilon'/k_BT)]^{1/(1-\gamma)} \qquad (8).$$

This also gives for the isothermal annealing curves at a given temperature $T_a$ versus time according to the following equations for the $\gamma^{th}$-order reaction kinetics:

$$N/N_0 = (1 + \alpha t)^{1/(1-\gamma)} \qquad (9),$$

with:

$$\alpha = \nu_0' N_0^{\gamma-1} (\gamma-1) \exp(-\Delta\varepsilon'/k_BT_a) \qquad (10).$$

The second-order reaction kinetics ($\gamma=2$) is classically used for Frenkel pair recombination [11]. It is to be noted that the decay time ($\alpha^{-1}$) is in this case dependent of $N_0$.

The least-square fitted activation energy ($\Delta\varepsilon'$) obtained by using Eq(8) with $\gamma=2$ is 0.76 eV for isochronal annealing of $F^+$-like centres in sample AU-11 (Fig. 3), which is not very different from the values (0.70 eV) extracted from the first-order kinetics (Table 1). However, the isochronal annealing curves of $F^+$-like centres in the electron-irradiated samples cannot be satisfactorily fitted with such a temperature-dependence. The annealing curves are always too steep to match the smoother decrease of second-order kinetics. Moreover, the isothermal annealing curves of lines 1 or 2 cannot also be properly fitted by the time-dependence given by Eq(9) with $\gamma=2$ versus reduced annealing time ($t/\tau$) (Fig. 4).

A similar deviation from the first-order exponential decay on a much larger time scale ($t\leq 200$ h) was already found for the isothermal annealing curve of $F^+$-like centres at 295 K (RT ageing) in a 2.5-MeV electron irradiated sample at $9.8\times10^{17}$ cm$^{-2}$, with a larger $N_0$ value, even if the isochronal annealing curve was satisfactorily fitted by a first-order reaction kinetics [4]. For ageing times up to 50 h ($t<2\tau$), the time decay is fitted by Eq(5) with $\tau=26$ h [4] and plotted versus $t/\tau$ (Fig. 4). At longer times, we found an asymptotic behaviour similar to the 2.0-MeV electron irradiated samples (q, r, t), which cannot also be fitted by a second-order ($\gamma=2$) decay (Fig. 4). All these data cannot be properly fitted by the third-order kinetics model ($\gamma=3$) either (Fig. 4).

The asymptotic values of all electron-irradiated samples are almost equal for $F^+$-like centres ($N_\infty\approx 1.5\times10^{16}$ cm$^{-3}$), despite different $N_0$ values. Moreover, a similar normalized asymptotic behaviour is found for the hole centres in sample q with $N_\infty/N_0 \approx 0.4$ (Fig. 4) and a $N_0$ value smaller by a factor about 4, and with a defect life time $\tau=1.4h$ near that of $F^+$-like centres at 394 K (Table 1).

This means that the kinetics of type I recovery process of $F^+$-like centres is certainly more complex than these crude approximations. This process could be correlated with the hole centre recovery and governed by electron ($V_O^-$) and hole ($O^-$) recombination through tunnelling between localized levels in the wide band gap of the insulator [13]. Nevertheless, consistent activation energies can be obtained on the basis of these simple models. The Arrhenius plot of $\tau$ versus $1/T_a$, according to Eq(6) yields then $\Delta\varepsilon=0.38$ eV (Fig. 5). This is in rather good agreement with the activation energies deduced with the first-order kinetics from the isochronal annealing curves for an annealing time (0.5h) smaller than $\tau$ (Table 1: samples j and s).

### IV.2. Type II recovery process

Type II process exhibits two-stages: in the first stage below 550 K, similar $\Delta\varepsilon$ and $T_{1/2}$ values were previously found for the $F^+$-like centres as in type I [4]. In the second stage above 550 K, new lines appeared and caused apparent increase of $F^+$-type centres in the case of 2.5-MeV electron or 100-MeV C ion irradiations [4].

A clear correlation of this second stage is seen with T-centres. Indeed, type II occurs only if T-centres are still not annealed out at 550 K, and if their concentration is large enough. Such a recovery process does not occur in the Au and I ion irradiated samples with a weak line 3, or in sample j, in which the T-centre concentration is also low ($2\times10^{18}$ cm$^{-3}$) even if $T_{1/2}=870$ K.

It thus depends on fluence, since T-centres are far more stable in as-received samples irradiated at high fluences with electrons or ions ($T_{1/2} \approx 900$ K) [3-5]. It also depends on the TCR treatment at 1373 K under vacuum prior to the irradiations, since T-centres are bleached at about 950 K ($T_{1/2} \approx 900$ K) in these reduced samples [4]. It is interesting to compare the as-received and reduced samples irradiated with C ions in the previous data [4]: type II indeed occurs in the reduced samples (e.g. sample C-20), but not in the as-received ones (e.g. sample C-19) irradiated exactly at the same fluence ($9.3\times10^{14}$ cm$^{-2}$). The T-centre concentration is larger in the former case ($\approx6\times10^{18}$ cm$^{-3}$) than in the latter one ($\approx2\times10^{18}$ cm$^{-3}$) [4]. The same concentration ratio of about 3 is found between as-received and reduced samples irradiated with C ions at a lower fluence ($1.0\times10^{14}$ cm$^{-2}$), and in as-received samples irradiated with 2.5-MeV electrons for a fluence increase by a factor around 10 [4]. Therefore, we conclude that T centres must not be annealed at 550K and reach a critical concentration in order to trigger the second stage of type II recovery process.

A new analysis of this second stage is made in the present work by fitting the spectra of isochronally annealed sample C-20 with four Lorentzian lines instead of the three ones previously used in ref [4], with a new line (6) growing at g=1.982 in between line 1 (g=1.997) and line 2 (g=1.973) (Fig. 1(b)). It is observed that lines 1 and 2 decrease gradually above 400 K, whereas line 6 start to grow. The densities of $F^+$-like centres corresponding to lines 1 and 2 (N), and paramagnetic centres corresponding to line 6 (N'), are normalized by the initial density of $F^+$-like centres ($N_0$), and plotted versus isochronal annealing temperature (Fig. 6).

In this case, Eq(1) must be rewritten as:

$$dN/dt = -\nu N - \nu_1 N \qquad (11),$$

where $\nu$ and $\nu_1$ are respectively the frequency factors for thermal annealing and conversion of $F^+$-like centres, assuming first-order reaction kinetics for both processes. Moreover, another rate equation must be added for the new centres:

$$dN'/dt = \nu_1 N - \nu_2 N' \qquad (12),$$

where $\nu_2$ is the frequency factor for thermal annealing of these centres, assuming also a first-order reaction kinetics.

Integration of Eq(11) yields:

$$N/N_0 = \exp[-(\nu + \nu_1) t] \qquad (13).$$

Assuming that $\nu_1/\nu$ is equal to a constant value K versus temperature and using Eq(2), Eq(13) yields then for the isochronal annealing of $F^+$-like centres:

$$N/N_0 = \exp[-(K+1)\nu_0 t_a \exp(-\Delta\varepsilon/k_B T)] \qquad (14),$$

which is similar to Eq(3) in the case of type I process.

The $F^+$-like centres are thus annealed during the first stage above 400 K, with $\Delta\varepsilon$=0.27 eV, according to a least-squares fit to Eq(14) (Fig. 6), in agreement with our previous value of 0.23 eV (for line 1 only) on the basis of Eq(3) [4]. This confirms our assumption that the $F^+$-like centre annealing can be modelled on the basis of only one activation energy. At the characteristic temperature $T_{1/2}$ about 500 K, in agreement with our previous value of 500 K also for line 1 only [4], half of $F^+$-like centres are annealed and converted to the new centres with a normalized density ($N'/N_0$) which exhibits a plateau at about 0.7 above 700 K (Fig. 6). Annealing of the latter centres starts above 750 K, and is almost completed at 973 K [4].

After some straightforward but lengthy algebra, one finds by integrating Eq(12):

$$N'/N_0 = [\nu_1 / (\nu + \nu_1 - \nu_2)] \{\exp(-\nu_2 t) - \exp[-(\nu + \nu_1)t]\} \quad (15),$$

Using a thermally-activated behaviour for the three frequency factors ($\nu$, $\nu_1$, $\nu_2$) like in Eq(2), and assuming that $\nu_2 \ll \nu + \nu_1$ (see below), Eq(15) is then approximated by a function of temperature of the kind:

$$N'/N_0 = K/(1+K) \{\exp[-A \exp(-\Delta\varepsilon_2/k_B T)] - \exp[-B \exp(-\Delta\varepsilon_1/k_B T)]\} \quad (16),$$

where A and B are positive constants, and $\Delta\varepsilon_1$ and $\Delta\varepsilon_2$ the activation energies for defect production and annealing respectively. Least-squares fit of the data to Eq(16) (Fig. 6) gives $K/(1+K) = 0.69$, $\Delta\varepsilon_1 = 0.40$ eV, and $\Delta\varepsilon_2 = 1.76$ eV. This yields a K value of around 2.23, meaning that during the second stage of the annealing process, about 70% of $F^+$-like centres are annealed out and 30% are converted to the new centres. It is seen that $\Delta\varepsilon$ and $\Delta\varepsilon_1$ are close values, in agreement with our assumption of that the new centres arise from the $F^+$-like centres. Moreover, it is to be noted that the calculated ratio $(\nu + \nu_1)/\nu_2$ is very large in this temperature range, thereby justifying also our approximation in Eq(16).

It is interesting to note that T centres are annealed above 650 K, with $\Delta\varepsilon = 0.57$ eV and $T_{1/2} = 890$ K [4], according to Eqs(3)-(4) (Fig. 6), in agreement with our previous values of 0.60 eV and 890 K [4], in the same temperature range as the new defects (Fig. 6). This shows again the correlation between these new paramagnetic centres and the T centres.

## V. CONCLUSIONS

We have studied the thermal recovery of the colour centres induced in cubic yttria-stabilized zirconia (YSZ) single crystals, i.e. $ZrO_2$:Y with 9.5 mol% $Y_2O_3$, 1.0- and 2.0-MeV electron, and 200-MeV $^{127}$I or 200-MeV $^{197}$Au ion irradiations: two electron centres namely, i) an $F^+$-like centre (singly ionized oxygen vacancy), ii) the so-called T-centre ($Zr^{3+}$ in a trigonal oxygen local environnement), also obtained by photon irradiations or thermo-chemical reduction (TCR) under vacuum, and iii) a hole centre. The defect stability is found to increase when fluence increases, or when thermo-chemical is carried out at 1373 K under vacuum, prior to the irradiations. Two distinct recovery processes are observed: i) type I is a simple single-stage thermally-activated process, with a non-zero limit value of the isothermal annealing curves of $F^+$-like centres for long annealing times, which might be linked to electron-hole recombination, ii) type II is a complex two-stage process with a first stage similar to type I below 550 K, and a second stage where transformation of the $F^+$-like centres into new defects occurs above 550 K. Simple kinetics models based on first-order reactions are proposed for both annealing processes. Single-stage processes occurs for $F^+$-like centres in as-received samples irradiated at low fluences, and for T-centres regardless of fluence. Type II is found only for $F^+$-like centres in as-received samples irradiated at high fluences, for larger defect concentrations, and in vacuum-annealed reduced samples, regardless of fluence. In those cases, a clear correlation is found between the type II recovery process and a critical concentration of T-centres stable above 550 K.

**ACKNOWLEDGMENTS:** Dr. S. Esnouf (LSI, Ecole Polytechnique, Palaiseau, France), Dr. J. J. Grob, Dr. J. P. Stoquert (PHASE Laboratory, Strasbourg), and Dr F. Hass (IReS, Strasbourg) are thanked for their kind help during the irradiations.

**Table 1:** Thermal annealing data of $F^+$-type centres (with volume densities $N_0$ corresponding to the sum of lines 1 and 2 intensities) induced in YSZ (100) single crystals after irradiation with charged particles of energy E, and mean particle range R computed with the ESTAR code [7] for electrons and the TRIM-96 code [8] for ions: activation energies for recovery ($\Delta\varepsilon$), frequency pre-factors ($\nu_0$), and temperatures ($T_{1/2}$) at which 50% of defects are isochronally annealed out, on the basis of first-order kinetics, activation energies for recovery ($\Delta\varepsilon'$), on the basis of second-order kinetics, and defect lifetime ($\tau$) during isothermal annealing at temperature $T_a$.

| Sample | Particle | E [MeV] | R [µm] | Fluence [cm$^{-2}$] | $N_0$ [cm$^{-3}$] | $\Delta\varepsilon$ [eV] | $\nu_0$ [s$^{-1}$] | $T_{1/2}$ [K] | $\Delta\varepsilon'$ [eV] | $T_a$ [K] | $\tau$ [h] |
|---|---|---|---|---|---|---|---|---|---|---|---|
| j | e$^-$ | 1.0 | 1.1x10$^3$ | 1.0x10$^{19}$ | 8.0x10$^{15}$ | 0.45(line2) | 93 | 436 | | | |
| q | e$^-$ | 2.0 | 2.4x10$^3$ | 3.2x10$^{18}$ | 3.8x10$^{16}$ | | | | | 394 | 0.8(line1) 1.1(line2) |
| r | e$^-$ | 2.0 | 2.4x10$^3$ | 3.2x10$^{18}$ | 3.6x10$^{16}$ | | | | | 368 | 1.8(line1) 1.3(line2) |
| s | e$^-$ | 2.0 | 2.4x10$^3$ | 3.2x10$^{18}$ | 3.5x10$^{16}$ | 0.59(line2) | 730 | 492 | | | |
| t | e$^-$ | 2.0 | 2.4x10$^3$ | 3.2x10$^{18}$ | 3.4x10$^{16}$ | | | | | 350 | 8.0(line1) |
| I-6 | $^{127}$I | 200 | 12 | 1.0x10$^{12}$ | 3.2x10$^{17}$ | 0.68(line2) | 2.0x10$^4$ | 456 | | | |
| I-14 | $^{127}$I | 200 | 12 | 1.0x10$^{13}$ | 6.7x10$^{17}$ | 0.33(line1) 0.33(line2) | 1.5 1.8 | 483 474 | | | |
| I-18 | $^{127}$I | 200 | 12 | 3.0x10$^{13}$ | 3.8x10$^{17}$ | 0.25(line1) 0.25(line2) | 0.13 0.14 | 540 515 | | | |
| AU-5 | $^{197}$Au | 200 | 10 | 1.0x10$^{12}$ | 4.9x10$^{17}$ | 0.66(line1) 0.69(line2) | 1.2x10$^4$ 1.2x10$^4$ | 449 452 | | | |
| AU-11 | $^{197}$Au | 200 | 10 | 2.5x10$^{13}$ | 2.8x10$^{17}$ | 0.70(line1) 0.83(line2) | 2.4x10$^3$ 1.6x10$^4$ | 532 562 | 0.76(line1) | | |

**Table 2:** Thermal annealing data of T-centres (same parameters as in Table 1, with volume densities $N_0$ corresponding to lines 3 intensities)).

| Sample reference | Particle | E [MeV] | R [μm] | Fluence [cm$^{-2}$] | $N_0$ [cm$^{-3}$] | Δε [eV] | $\nu_0$ [s$^{-1}$] | $T_{1/2}$ [K] |
|---|---|---|---|---|---|---|---|---|
| e | e$^-$ | 1.0 | 1.1x10$^3$ | 9.7x10$^{17}$ | 1.7x10$^{18}$ | 0.41 | 12.8 | 479 |
| j | e$^-$ | 1.0 | 1.1x10$^3$ | 1.0x10$^{19}$ | 2.0x10$^{18}$ | 1.28 | 1.6x10$^4$ | 870 |
| s | e$^-$ | 2.0 | 2.4x10$^3$ | 3.2x10$^{18}$ | 1.3x10$^{18}$ | 0.54 | 186 | 492 |

**Fig. 1:** Room-temperature EPR spectrum with B//<100> of as-received YSZ (100) single crystal irradiated with 2.0-MeV electrons (sample q) (a); and vacuum-annealed reduced YSZ (100) single crystal irradiated with 100-MeV C ions (sample C-20) and annealed in air at 873 K [4] (b).

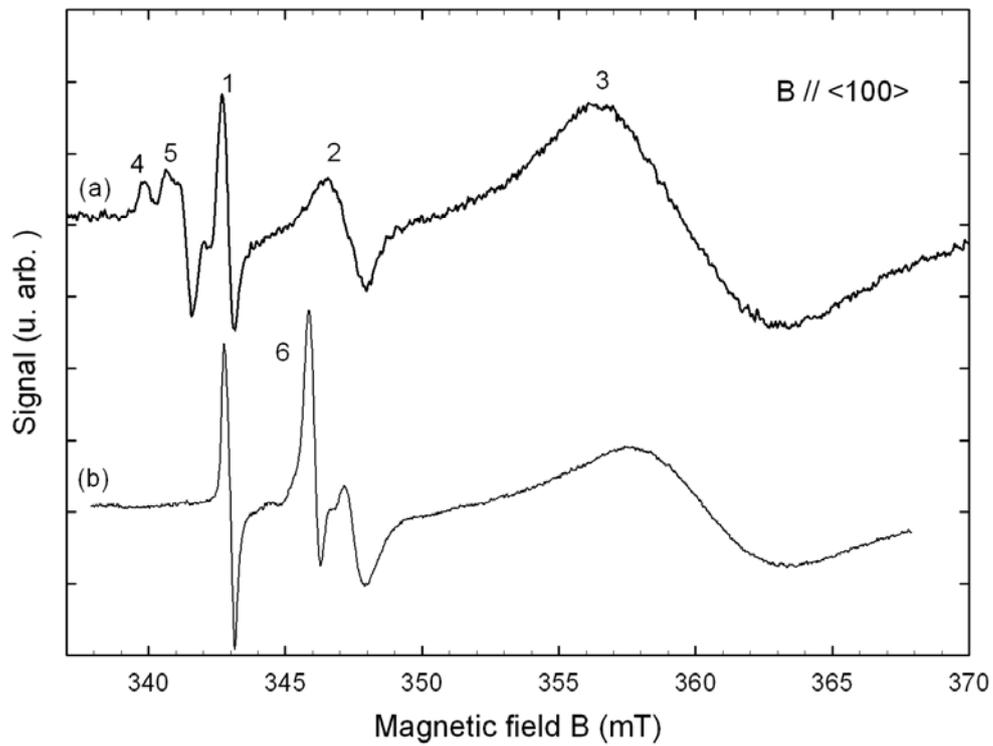

**Fig. 2:** Normalized paramagnetic centre density versus isochronal annealing temperature for electron- irradiated YSZ single crystals; sample e (1.0-MeV): T centres (line 3, g=1.904, open triangles); sample j (1.0-MeV): $F^+$-like centres (line 1, g=1.998, full squares), T centres (line 3, g=1.904, open squares), T centre satellite (g=1.97, full diamonds); sample s (2.0-MeV): $F^+$-like centres (line 2, g=1.973, full circles), T centres (line 3, g=1.904, open circles), hole centres (line 5, g=2.007, open diamonds). Lines are least-squares fits to the first-order kinetics.

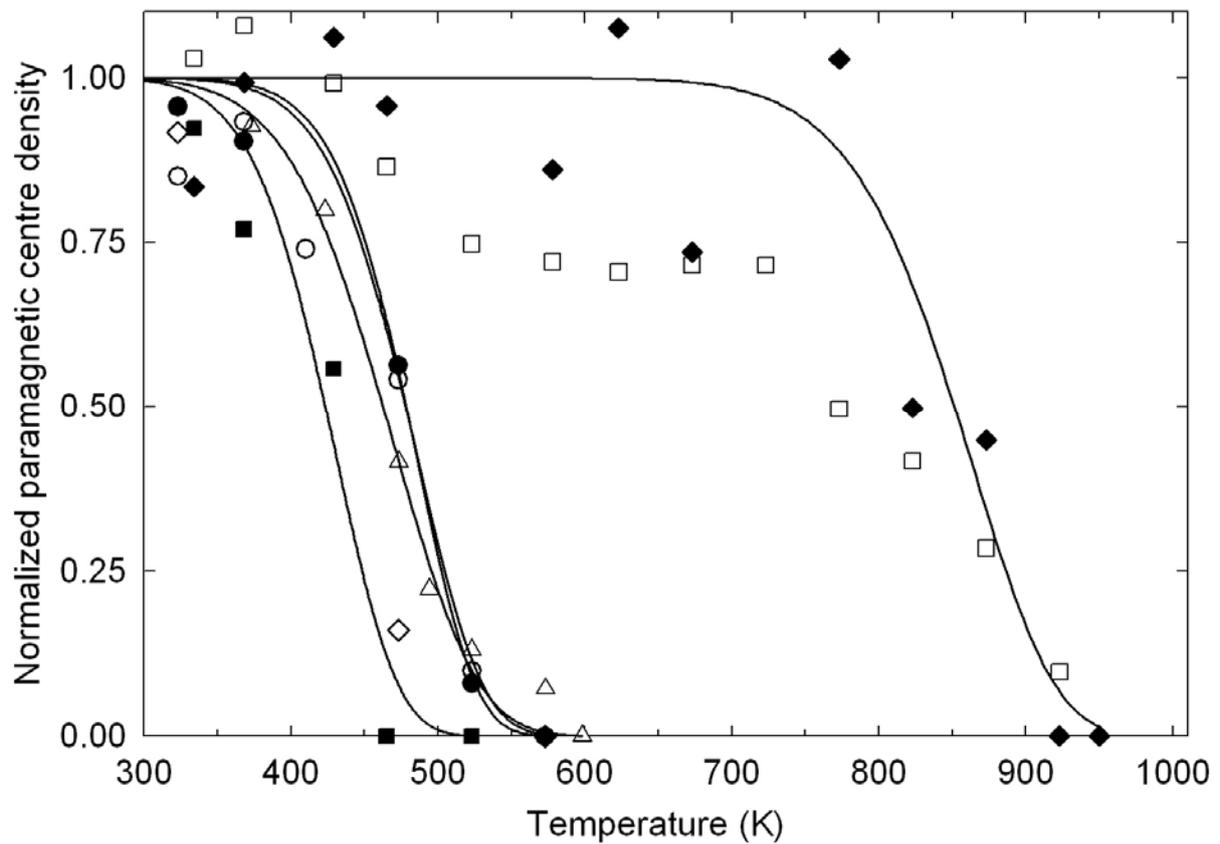

**Fig. 3:** Normalized paramagnetic F$^+$-like centre density versus isochronal annealing temperature for YSZ single crystals irradiated with 200-MeV I ions: sample I-14 (line 1, g=1.998, open circles); sample I-18 (line 1, g=1.998, full circles), and with 200-MeV Au ions: sample AU-5 (line 1, g=1.998, open squares), sample AU-11 (line 1, g=1.998, full squares). Solid lines are least-squares fits to the first-order kinetics. Dashed line is a least-squares fit to the second-order kinetics for AU-11 sample.

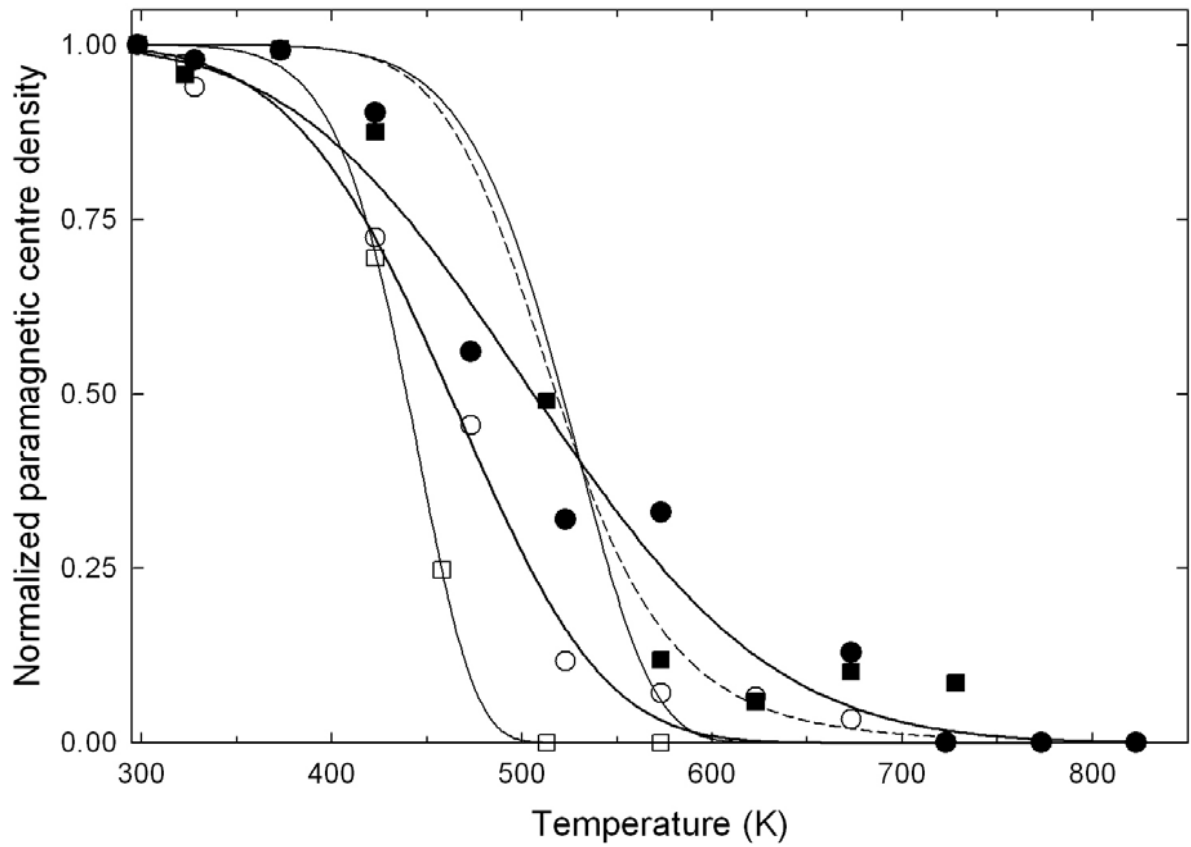

**Fig. 4:** Normalized paramagnetic centre density versus reduced annealing time for YSZ samples irradiated with 2.0-MeV electrons at the same fluence (3.2x10[18] cm[-2]) and annealed at 350 K (sample t: g=1.998, full down triangles), 368 K (sample r: line 1, g=1.998, full circles; line 2, g=1.973, full squares), and 394 K (sample q: line 1, g=1.998, open circles; line 2, g=1.973, open squares; line 5, g=2.007, open diamonds), and irradiated with 2.5-MeV electrons at 9.8x10[17] cm[-2] and aged at 300 K (sample f: open up triangles) [4]. The solid line is a least-squares fit for reduced times < 2. Dashed lines are least-squares fits to the second-order kinetics. Dash-dotted lines are least-squares fits to the third-order kinetics.

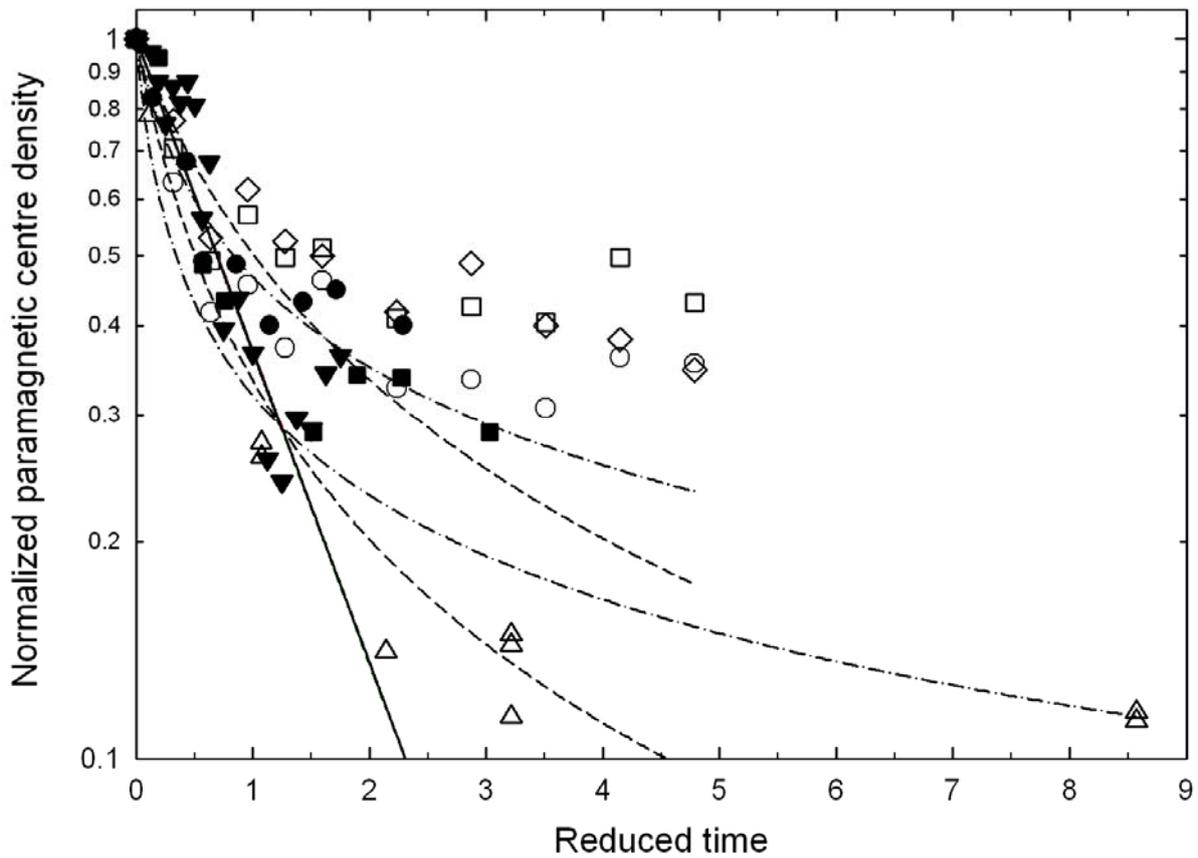

**Fig. 5:** Paramagnetic centre life time versus inverse isothermal annealing temperature for YSZ samples (q, r, t) irradiated with 2.0-MeV electrons at the same fluence ($3.2 \times 10^{18}$ cm$^{-2}$): F$^+$-like centres (open circles), hole centres (open square).

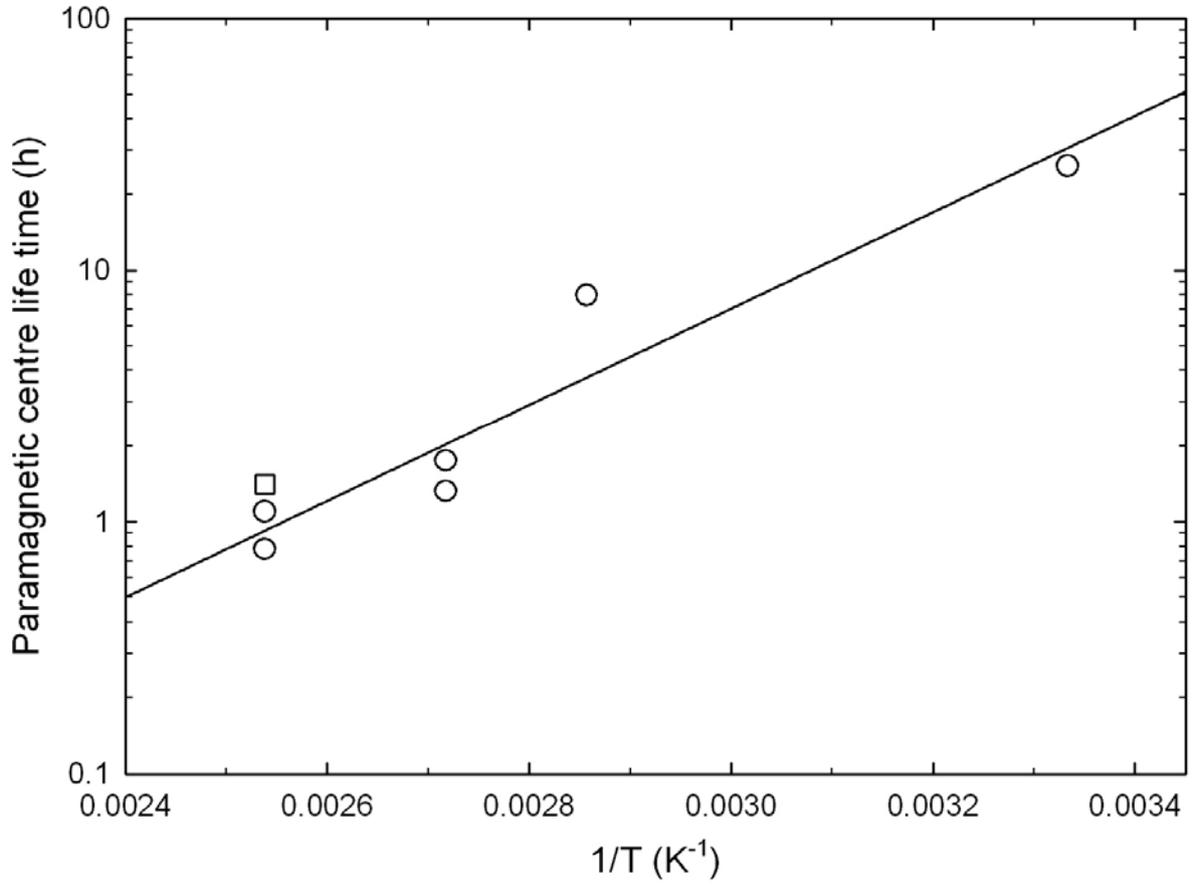

**Fig. 6:** Normalized paramagnetic centre density versus isochronal annealing temperature for an YSZ single crystal (sample C-20) irradiated with 100-MeV C ions at a fluence of $9.3 \times 10^{14}$ cm$^{-2}$ [4]: F$^+$-like centres (open circles), T centres (open squares), new centres (full triangles). Solid lines are least-squares fits to first-order kinetics modelling.

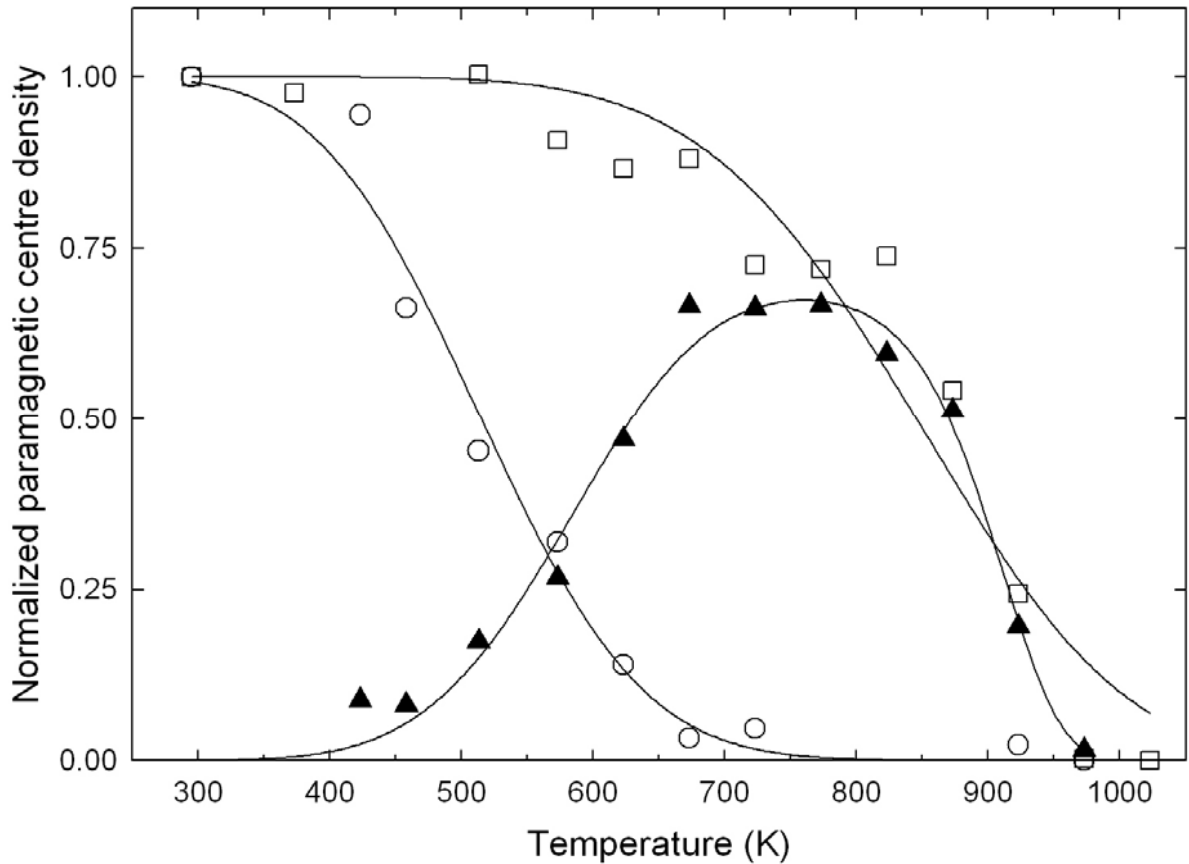